\documentclass[aps,twocolumn,pra,superscriptaddress,showpacs,tightenlines]{revtex4-1}
\usepackage [latin1]{inputenc}
\usepackage{amsmath}
\usepackage{graphicx}
\usepackage{color}
\usepackage{amsfonts}
\usepackage{txfonts}
\usepackage[colorlinks,citecolor=blue]{hyperref}
\begin{document}

\title{Domino cooling of a coupled mechanical-resonator chain via cold-damping feedback}

\author{Deng-Gao Lai}
\affiliation{Key Laboratory of Low-Dimensional Quantum Structures and Quantum Control of Ministry of Education, Key Laboratory for Matter Microstructure and Function of Hunan Province, Department of Physics and Synergetic Innovation Center for Quantum Effects and Applications, Hunan Normal University, Changsha 410081, China}
\affiliation{Theoretical Quantum Physics Laboratory, RIKEN Cluster for Pioneering Research, Wako-shi, Saitama 351-0198, Japan}

\author{Jian Huang}
\affiliation{Key Laboratory of Low-Dimensional Quantum Structures and Quantum Control of Ministry of Education, Key Laboratory for Matter Microstructure and Function of Hunan Province, Department of Physics and Synergetic Innovation Center for Quantum Effects and Applications, Hunan Normal University, Changsha 410081, China}

\author{Bang-Pin Hou}
\affiliation{College of Physics and Electronic Engineering, Institute of Solid State Physics, Sichuan Normal University, Chengdu 610068, China}

\author{Franco Nori}
\affiliation{Theoretical Quantum Physics Laboratory, RIKEN Cluster for Pioneering Research, Wako-shi, Saitama 351-0198, Japan}
\affiliation{Physics Department, The University of Michigan, Ann Arbor, Michigan 48109-1040, USA}

\author{Jie-Qiao Liao}
\email{jqliao@hunnu.edu.cn}
\affiliation{Key Laboratory of Low-Dimensional Quantum Structures and Quantum Control of Ministry of Education, Key Laboratory for Matter Microstructure and Function of Hunan Province, Department of Physics and Synergetic Innovation Center for Quantum Effects and Applications, Hunan Normal University, Changsha 410081, China}

\begin{abstract}
We propose a domino-cooling method to realize simultaneous ground-state cooling of a coupled mechanical-resonator chain through an optomechanical cavity working in the unresolved-sideband regime. This domino-effect cooling is realized by combining the cold-damping feedback on the first mechanical resonator with nearest-neighbor couplings between other neighboring mechanical resonators. We obtain analytical results for the effective susceptibilities, noise spectra, final mean phonon numbers, and cooling rates of these mechanical resonators, and find the optimal-cooling condition for these resonators. Particularly, we analyze a two-mechanical-resonator case and find that by appropriately engineering either the laser power or the feedback, a flexible switch between symmetric and asymmetric ground-state cooling can be achieved. This could be used for preparing symmetric quantum states in mechanical systems. We also simulate the cooling performance of a coupled $N$-mechanical-resonator chain and confirm that these resonators can be simultaneously cooled to their quantum ground states in the unresolved-sideband regime. Under proper parameter conditions, the cooling of the mechanical-resonator chain shows a temperature gradient along the chain. This study opens a route to quantum manipulation of multiple mechanical resonators in the bad-cavity regime.
\end{abstract}

\maketitle

\section{Introduction\label{sec1}}

Cavity optomechanical systems~\cite{Kippenberg2008Science,Meystre2013AP,Aspelmeyer2014RMP}, addressing the radiation-pressure coupling between mechanical motion of mesoscopic or even macroscopic objects and electromagnetic degrees of freedom, provide a promising platform for manipulating cavity-field statistics by mechanically changing the cavity boundary or controlling the mechanical properties through optical means ~\cite{Rabl2011PRL,Nunnenkamp2011,Liao2012PRA,Liao2013PRA,Liao2013,Wang2013PRL,Liu2013PRL,Vitali2007PRL,Agarwal2010PRA,Genes2011PRA,Cirio2017PRL,Xu12015PRA,Hou2015PRA,Wu2018PRApplied,Qin2018PRL,Zippilli2018PRA}. Optomechanical cooling~\cite{Wilson-Rae2007PRL,Marquardt2007PRL,Genes2008NJP,Xia2009PRL,Liu2013PRL1,Xu2017PRL,Teufel2011Nature,Clarkl2017Nature,MXu2020PRL,Qiu2020PRL,Mancini1998PRL,Genes2008PRA,Steixner2005PRA,Bushev2006PRL,Rossi2017PRL,Rossi2018Nature,Conangla2019PRL,Tebbenjohanns2019PRL,Sommer2019PRL,Guo2019PRL,Sommer2020PRR}, as a prominent application closely relevant to this platform, has become an important research topic in this field. This is because a prerequisite for observing the signature of quantum mechanical effects is to cool the systems to their quantum ground states, such that thermal noise can be suppressed. So far, the ground-state cooling of a single mechanical resonator based on optomechanical platforms has been mainly achieved by two cooling mechanisms: (i) resolved-sideband cooling~\cite{Wilson-Rae2007PRL,Marquardt2007PRL,Genes2008NJP,Xia2009PRL,Liu2013PRL1,Xu2017PRL,Teufel2011Nature,Clarkl2017Nature,MXu2020PRL,Qiu2020PRL}, which is preferable in the good-cavity regime; and (ii) feedback-aided cooling~\cite{Mancini1998PRL,Genes2008PRA,Steixner2005PRA,Bushev2006PRL,Rossi2017PRL,Rossi2018Nature,Conangla2019PRL,Tebbenjohanns2019PRL,Sommer2019PRL,Guo2019PRL,Sommer2020PRR}, which is more efficient in the bad-cavity regime. Alternatively, cooling can also be achieved in superconducting quantum circuits~\cite{Grajcar2008PRB,Zhang2009PRA,Liberato2011PRA,Xue2007PRB,You2008PRL,Nori2008NP,Xiang2013RMP}. Note that the ground-state cooling of the mechanical resonators means that the final average occupancies in these resonators are well below unity~\cite{Wilson-Rae2007PRL,Marquardt2007PRL}.

In recent years, much attention has been paid to the multimode optomechanical systems involving multiple mechanical resonators~\cite{Shkarin2014PRL,Malz2018PRL,Shen2016NP,Shen2018NC,Fang2017NP,Xu2019Nature,Mathew2018arXiv,Yang2020NC,Massel2012Nc,Mari2013PRL,Matheny2014PRL,Zhang2015PRL,Riedinger2018Nature,Ockeloen-Korppi2018,Stefano2019PRL,Li2020PRA,Pelka2020PRR,Xuereb2012PRL,Xuereb2014PRL,Heinrich2011PRL,Ludwig2013PRL,Xuereb2015NJP,Mahmoodian2018PRL,Xu2016Nature,SanavioPRB2020}. The motivations for exploring these systems include the study of macroscopic mechanical coherence in multimode mechanical systems~\cite{Shkarin2014PRL,Massel2012Nc,Mari2013PRL,Matheny2014PRL,Zhang2015PRL,Riedinger2018Nature,Riedinger2018Nature,Ockeloen-Korppi2018,Stefano2019PRL,Li2020PRA,Pelka2020PRR}, the engineering of complex long-range interactions among the mechanical components~\cite{Xuereb2012PRL,Xuereb2014PRL}, the investigation of quantum many-body phenomena~\cite{Heinrich2011PRL,Ludwig2013PRL,Xuereb2015NJP,Mahmoodian2018PRL}, and the implementation of nonreciprocal photon or phonon transport~\cite{Malz2018PRL,Shen2016NP,Shen2018NC,Fang2017NP,Xu2019Nature,Mathew2018arXiv,Yang2020NC,Xu2016Nature,SanavioPRB2020}. However, these applications are fundamentally limited by thermal noise. To suppress these thermal effects, the simultaneous ground-state refrigeration of these mechanical resonators becomes an urgent and important task. Although some schemes for cooling multiple mechanical resonators in the good-cavity regime have been proposed using the cavity resolved-sideband-cooling mechanism~\cite{Ockeloen-Korppi2019PRA,Lai2018PRA,Lai2020PRARC}, the answer to the question whether we can utilize the feedback-cooling technique to simultaneously cool these mechanical resonators to their quantum ground states is yet unclear.

\begin{figure*}[tbp]
\center
\includegraphics[width=1 \textwidth]{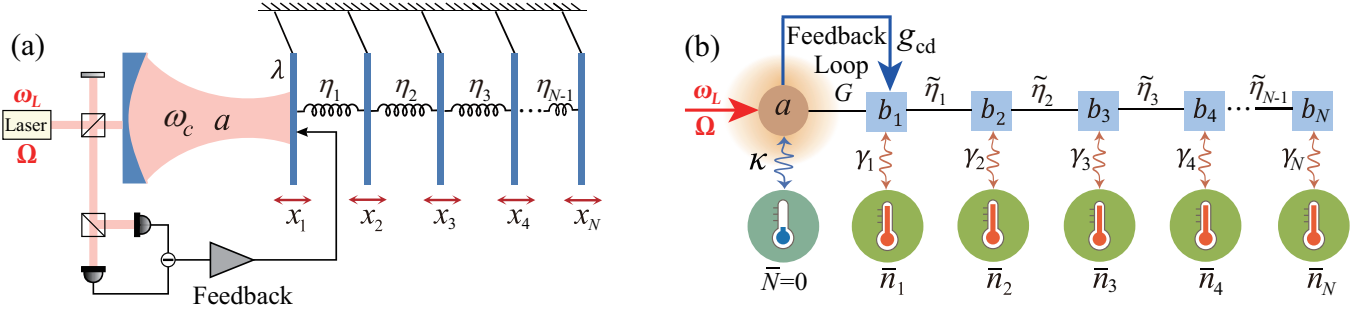}
\caption{(a) Schematic of the cascade optomechanical system. A cavity field with resonance frequency $\omega_{c}$ couples to the first mechanical resonator in a coupled mechanical-resonator chain via radiation-pressure coupling with coupling strength $\lambda$. Other neighboring mechanical resonators are coupled to each other through a position-position coupling with strength $\eta_{j}$ for $j=1$-$(N-1)$. A monochromatic laser with frequency $\omega_{L}$ and field amplitude $\Omega_{L}$ is introduced to drive the cavity. The output field of the driven optomechanical cavity is measured via the homodyne detection. A feedback loop is utilized to design a direct force applied upon the first resonator, which can lead to the freezing of its thermal fluctuations (cold-damping effect). (b) By performing a linearization, the model (a) can be simplified to a cascade-coupled bosonic system where a cavity-field mode $a$ couples to an array of $N$ mechanical modes connected in series. The entire chain is cooled via a domino effect or chain reaction through the system. The nearest-neighboring mechanical modes are coupled to each other through an effective nearest-neighbor coupling with strength $\tilde{\eta}_{j}$ for $j=1$-$(N-1)$. The cavity-field mode $a$ is coupled to a vacuum bath ($\bar{N}=0$ and the decay rate $\kappa$), and each mechanical mode is connected to its own heat bath (with thermal phonon numbers $\bar{n}_{j}$ and decay rates $\gamma_{j}$). The feedback loop applied on the first mechanical mode $b_{1}$ is characterized by the linearized optomechanical coupling $G$ and the feedback gain $g_{\text{cd}}$.}
\label{Figmodel}
\end{figure*}

In this paper, we demonstrate that an array of $N$ mechanical resonators coupled in series can be simultaneously cooled to their quantum ground states with cold-damping feedback. Here, the feedback technique is applied to the optomechanical cavity via a feedback loop, which is utilized to design a direct force applied on the first resonator. This leads to the freezing of thermal fluctuations of the first mechanical resonator (cold-damping effect). Other neighboring mechanical resonators are connected to each other via position-position interactions (namely nearest-neighbor interactions). Physically, the feedback loop applied on the first mechanical resonator acts as a cooling channel of the first mechanical resonator. Successively, the former resonator provides a cooling channel for the next resonator via the nearest-neighbor coupling, as a cascade-cooling process. This acts like a domino-effect or chain-reaction cooling through the system.

By deriving analytical results of the effective susceptibilities, noise spectra, final mean phonon numbers, and cooling rates of these mechanical resonators, we obtain the optimal-cooling condition for this coupled mechanical-resonator chain. Our proposal allows both degenerate and non-degenerate mechanical resonators to reach simultaneous ground-state cooling in the unresolved-sideband regime. We also find that a flexible switch between asymmetric and symmetric ground-state coolings can be achieved by appropriately engineering either the laser power or the feedback parameters (e.g., feedback gain and feedback bandwidth) applied on the first mechanical resonator. Note that the asymmetric (symmetric) cooling means that the final mean phonon numbers of the two mechanical resonators are different (the same). The symmetric cooling case will be helpful to the creation of symmetric quantum states in the two mechanical resonators, because their initial states are almost the same.

Additionally, we extend this domino cooling method to the simultaneous cooling of $N$ mechanical resonators. The results show that, when the mechanical coupling strength is much smaller than the mechanical frequency, the cooling efficiency is higher for the mechanical resonator which is closer to the cavity. Physically, the feedback loop extracts the thermal excitations from the first resonator though the feedback cooling channel, and then the feedback-cooled resonator extracts the thermal excitations from the next one via the mechanical cooling channel. In this case, the feedback cooling rate should be much larger than the mechanical cooling rates, which leads to the highest cooling efficiency for the feedback-cooled resonator. However, by increasing the mechanical coupling, an anomalous cooling occurs, i.e., the feedback-cooled resonator is not the coldest. This is because the counter-rotating-wave (CRW) interaction terms will create more and more phonon excitations with the increase of the mechanical coupling strength, and then the cooling of the first resonator is suppressed. This study will pave a way towards quantum manipulation of multimode mechanical systems in the bad-cavity regime.

The rest of this paper is organized as follows. In Sec.~\ref{sec2}, we introduce the physical model and the Hamiltonians. In Sec.~\ref{sec3}, we derive the Langevin equations and the final mean phonon numbers. In Secs.~\ref{sec4} and ~\ref{sec5}, we study the cooling of two and $N$ coupled mechanical resonators, respectively. Finally, we provide a brief conclusion in Sec.~\ref{sec6}. An Appendix is presented to display the detailed calculation of the final mean phonon numbers in the two-mechanical-resonator case.

\section{Model and Hamiltonian\label{sec2}}

We consider a multimode optomechanical system in which a single-mode cavity field couples to an array of $N$ mechanical resonators coupled in series, as illustrated in Fig.~\ref{Figmodel}(a). The first mechanical resonator is coupled to the cavity field via the radiation-pressure coupling, and these nearest-neighboring mechanical resonators are coupled to each other through position-position couplings (forming a cascade configuration). A strong driving field (the driving amplitude $\Omega$ is much larger than the cavity-field decay rate $\kappa$) is applied to the optical cavity for manipulating the optical and mechanical degrees of freedom. The Hamiltonian of the system reads ($\hbar =1$)~\cite{Lai2018PRA}
\begin{eqnarray}
H &=&\omega _{c}a^{\dagger }a+\sum_{j=1}^{N}\left(\frac{p_{x,j}^{2}}{2m_{j}}+\frac{m_{j}\tilde{\omega}_{j}^{2}x_{j}^{2}}{2}\right)-\lambda a^{\dagger}ax_{1} \notag \\
&&+\sum_{j=1}^{N-1}\eta_{j}(x_{j}-x_{j+1})^{2}+\Omega(a^{\dagger }e^{-i\omega _{L}t}+ae^{i\omega _{L}t}),\label{eq1iniH}
\end{eqnarray}
where $a$ ($a^{\dagger }$) is the annihilation (creation) operator of the cavity-field mode with the resonance frequency $\omega_{c}$. The momentum and position operators $p_{x,j}$ and $x_{j}$ describe the $j$th mechanical resonator with resonance frequency $\tilde{\omega}_{j}$ and mass $m_{j}$. The $\lambda$ term in Eq.~(\ref{eq1iniH}) denotes the optomechanical interaction between the cavity field and the first mechanical resonator, where $\lambda=\omega_{c}/L$ is the radiation-pressure force of a single photon, with $L$ being the rest length of the optical cavity.
The nearest-neighbor interactions between these neighboring mechanical resonators are depicted by these $\eta_{j}$ terms. The last term in Eq.~(\ref{eq1iniH}) describes the input laser driving with the driving frequency $\omega_{L}$ and amplitude $\Omega=\sqrt{2P_{L}\kappa/\omega_{L}}$, where $P_{L}$ and $\kappa$ are, respectively, the driving power and cavity-field decay rate.

For convenience, we introduce the dimensionless coordinator and momentum operators $q_{j}=\sqrt{m_{j}\omega_{j}}x_{j}$ and $p_{j}=\sqrt{1/(m_{j}\omega_{j})}p_{x,j}$ ($[q_{j},p_{j}]=i$) for $j\in[1,N]$, and the normalized resonance frequencies $\omega _{1(N)}=\sqrt{\tilde{\omega}_{1(N)}^{2}+2\eta_{1(N-1)}/m_{1(N)}}$ and $\omega_{j\in[2,N-1]}=\sqrt{\tilde{\omega}_{j}^{2}+2(\eta _{j-1}+\eta
_{j}) /m_{j}}$ for these resonators. In a rotating frame defined by the unitary transformation operator $\exp(-i\omega_{L}ta^{\dagger}a)$, Hamiltonian~(\ref{eq1iniH}) becomes
\begin{eqnarray}
H_{I} &=&\Delta _{c}a^{\dagger }a+\sum_{j=1}^{N}\frac{\omega _{j}}{2}\left(p_{j}^{2}+q_{j}^{2}\right) -\tilde{\lambda}a^{\dagger }aq_{1}  \notag \\
&&-\sum_{j=1}^{N-1}2\tilde{\eta}_{j}q_{j}q_{j+1}+\Omega (a^{\dagger}+a),\label{Hamlt2dimless}
\end{eqnarray}
where $\Delta_{c}=\omega_{c}-\omega_{L}$ is the driving detuning of the cavity field, $\tilde{\lambda}=\lambda \sqrt{1/(m_{1}\omega _{1})}$ and $\tilde{\eta}_{j\in[1,N-1]}=\eta _{j}\sqrt{1/(m_{j}m_{j+1}\omega _{j}\omega
_{j+1})}$ are, respectively, the strength of the optomechanical coupling and the mechanical interaction expressed with dimensionless coordinator and momentum operators.

\section{Langevin equations and final mean phonon numbers \label{sec3}}

In this section, we derive the quantum Langevin equations of the system, analyze the cold-damping feedback scheme, and obtain the final mean phonon numbers of the $N$-mechanical-resonator chain.

\subsection{Langevin equations\label{sec3A}}

To include the damping and noise effects in this system, we consider the case where the optical mode is coupled to a vacuum bath and $N$ mechanical modes are subjected to quantum Brownian forces. In this case, the evolution of the system can be described by the quantum Langevin equations
\begin{subequations}
\label{Langevineqorig}
\begin{align}
\dot{a}=&-[\kappa +i(\Delta_{c}-\tilde{\lambda} q_{1})]a-i\Omega +\sqrt{2\kappa }a_{\text{in}}, \\
\dot{q}_{j\in[1,N]}=&\omega _{j}p_{j}, \\
\dot{p}_{1}=&-\omega _{1}q_{1}+\lambda _{0}a^{\dagger }a+2\tilde{\eta}_{1}q_{2}-\gamma _{1}p_{1}+\xi_{1},  \\
\dot{p}_{j\in[2,N-1]}=&-\omega _{j}q_{j}+2\tilde{\eta}_{j-1}q_{j-1}+2\tilde{\eta}_{j}q_{j+1}-\gamma_{j}p_{j}+\xi _{j}, \\
\dot{p}_{N}=&-\omega _{N}q_{N}+2\tilde{\eta}_{N-1}q_{N-1}-\gamma_{N}p_{N}+\xi_{N},
\end{align}
\end{subequations}
where $\kappa$ and $\gamma_{j\in[1,N]}$ are, respectively, the decay rates of the cavity mode and the $j$th mechanical resonator. The operators $a_{\textrm{in}}$ $(a^{\dagger}_{\textrm{in}})$ and $\xi_{j\in[1,N]}$ denote the noise operators of the cavity field and the Brownian force acting on the $j$th mechanical resonator, respectively. These noise operators have zero mean values and the following correlation functions,
\begin{subequations}
\label{correlationfun}
\begin{align}
\langle a_{\textrm{in}}(t) a_{\textrm{in}}^{\dagger}(t^{\prime})\rangle=&\delta(t-t^{\prime}), \hspace{0.5 cm}
\langle a_{\textrm{in}}^{\dagger}(t) a_{\textrm{in}}(t^{\prime})\rangle =0, \\
\langle \xi_{j}(t)\xi_{j}(t^{\prime})\rangle=&\frac{\gamma_{j}}{\omega_{j}}\int \frac{d\omega }{2\pi}e^{-i\omega(t-t^{\prime})}\omega \left[\coth\left(\frac{\omega}{2k_{B}T_{j}}\right) +1\right],
\end{align}
\end{subequations}
where $k_{B}$ is the Boltzmann constant, and $T_{j\in[1,N]}$ is the temperature of the thermal reservoir associated with the $j$th mechanical resonator.

For cooling these mechanical resonators, the strong-driving regime of the cavity is considered, so that the average photon number in the cavity is sufficiently large and then we can simplify this physical model by a linearization procedure. To this end, we write the operators in Eq.~(\ref{Langevineqorig}) as sums of averages plus fluctuations: $o=\left\langle o\right\rangle_{\textrm{ss}} +\delta o$
for operators $a$, $a^{\dagger}$, $q_{j\in[1,N]}$, and $p_{j\in[1,N]}$. By separating the classical motion and quantum fluctuations, the linearized quantum Langevin equations become
\begin{subequations}
\label{fluceq}
\begin{align}
\delta \dot{X}=&-\kappa \delta X+\Delta \delta Y+\sqrt{2\kappa }X_{\text{in}}, \\
\delta \dot{Y}=&-\kappa \delta Y-\Delta \delta X+G\delta q_{1}+\sqrt{2\kappa }Y_{\text{in}}, \\
\delta \dot{q}_{j\in[1,N]}=&\omega _{j}\delta p_{j}, \\
\delta \dot{p}_{1}=&-\omega _{1}\delta q_{1}+G\delta X+2\tilde{\eta}_{1}\delta q_{2}-\gamma _{1}\delta p_{1}+\xi _{1}, \\
\delta \dot{p}_{j\in[2,N-1]}=&-\omega _{j}\delta q_{j}+2\tilde{\eta}_{j-1}\delta q_{j-1}+2\tilde{\eta}_{j}\delta q_{j+1}-\gamma _{j}\delta p_{j}+\xi _{j}, \\
\delta \dot{p}_{N}=&-\omega _{N}\delta q_{N}+2\tilde{\eta}_{N-1}\delta q_{N-1}-\gamma _{N}\delta p_{N}+\xi _{N},
\end{align}
\end{subequations}
where $X=(\delta a^{\dagger}+\delta a)/\sqrt{2}$ and $Y=i(\delta a^{\dagger}-\delta a)/\sqrt{2}$ are the quadratures of the cavity field, and $X_{\text{in}}$ and $Y_{\text{in}}$ denote the corresponding Hermitian input noise quadratures. Note that we have chosen the phase reference of the cavity field such that $\left\langle a\right\rangle_{\textrm{ss}}$ is real and positive. We have also defined the normalized driving detuning $\Delta=\Delta_{c}-\tilde{\lambda}\langle q_{1}\rangle_{\textrm{ss}}$ and the effective optomechanical coupling $G=\sqrt{2}\tilde{\lambda}\langle a\rangle_{\textrm{ss}}$ with $\langle a\rangle_{\textrm{ss}}=-i\Omega/(\kappa +i\Delta)$.

\subsection{Cold-damping feedback\label{sec3B}}

To realize the cold-damping feedback, we consider the case of $\Delta=0$, which indicates the highest sensitivity for position measurements of the mechanical resonator~\cite{Genes2008PRA,Sommer2019PRL}. Owing to the application of a negative derivative feedback, this cold-damping feedback technique can significantly increase the effective decay rate of the mechanical resonator without increasing the thermal noise~\cite{Courty2001EPJD,Vitali20024PRA}.

The position of the first mechanical resonator is measured through a phase-sensitive detection of the cavity output field, and then the readout of the cavity output field is fed back onto the first mechanical resonator by applying a feedback force. The intensity of the feedback force is proportional to the time derivative of the output signal, and therefore to the velocity of the first mechanical resonator~\cite{Courty2001EPJD,Vitali20024PRA,Genes2008PRA,Sommer2019PRL,Sommer2020PRR}. Then, the linearized quantum Langevin equations become
\begin{subequations}
\label{fluceqcd}
\begin{align}
\delta \dot{X}=&-\kappa \delta X+\sqrt{2\kappa }X_{\text{in}}, \\
\delta \dot{Y}=&-\kappa \delta Y+G\delta q_{1}+\sqrt{2\kappa }Y_{\text{in}}, \\
\delta \dot{q}_{j\in[1,N]}=&\omega_{j}\delta p_{j}, \\
\delta \dot{p}_{1}=&-\omega_{1}\delta q_{1}+G\delta X+2\tilde{\eta}_{1}\delta q_{2}-\gamma_{1}\delta p_{1}+\xi_{1}\notag \\
&-\int_{-\infty }^{t}g(t-s)\delta Y^{\text{est}}(s)ds, \\
\delta \dot{p}_{j\in[2,N-1]}=&-\omega _{j}\delta q_{j}+2\tilde{\eta}_{j-1}\delta q_{j-1}+2\tilde{\eta}_{j}\delta q_{j+1}-\gamma _{j}\delta p_{j}+\xi_{j}, \\
\delta \dot{p}_{N}=&-\omega_{N}\delta q_{N}+2\tilde{\eta}_{N-1}\delta q_{N-1}-\gamma_{N}\delta p_{N}+\xi_{N}.
\end{align}
\end{subequations}
In Eq.~(\ref{fluceqcd}d), the convolution term $\int_{-\infty }^{t}g(t-s)\delta Y^{\text{est}}(s)ds$ denotes the feedback force acting on the first mechanical resonator. This force depends on the past dynamics of the detected quadrature $\delta Y$, which is driven by the weighted sum of the fluctuations of the first mechanical resonator. The causal kernel is defined by~\cite{Genes2008PRA,Sommer2019PRL,Sommer2020PRR}
\begin{align}
&g(t)=g_{\text{cd}}\frac{d}{dt}[\theta(t)\omega_{\text{fb}}e^{-\omega_{\text{fb}}t}],
\end{align}
where $g_{\text{cd}}$ and $\omega_{\text{fb}}$ are the dimensionless feedback gain and the feedback bandwidth, respectively. The estimated intracavity phase quadrature $\delta Y^{\text{est}}$ results from the measurement of the output
quadrature $Y^{\text{out}}(t)$, which satisfies the usual input-output relation $\delta Y^{\text{out}}(t)=\sqrt{2\kappa}\delta Y(t)-Y_{\text{in}}(t)$. This relation is generalized to the case of a nonunit detection efficiency by modeling a detector with quantum efficiency $\zeta$ with an ideal detector preceded by a beam splitter (with transmissivity $\sqrt{\zeta}$), which mixes the incident field with an uncorrelated vacuum field $Y^{\upsilon}(t)$. Then, the estimated phase quadrature $\delta Y^{\text{est}}(t)$ is obtained as~\cite{Genes2008PRA,Sommer2019PRL,Sommer2020PRR}
\begin{align}
&\delta Y^{\text{est}}(t)=\delta Y(t)-\frac{Y_{\text{in}}(t)+\sqrt{\zeta^{-1}-1}Y^{\upsilon}(t)}{\sqrt{2\kappa}}.
\end{align}

Below, we seek for the steady-state solution of Eq.~(\ref{fluceqcd}) by solving the variables in the frequency domain with the Fourier transformation. We define the Fourier transform for an operator $r(t)=(1/2\pi)^{1/2}\int_{-\infty }^{\infty }e^{-i\omega t}\tilde{r}(\omega) d\omega$ ($r=\delta X$, $\delta Y$, $\delta q_{j}$, $\delta p_{j}$, $\xi_{j}$, $X_{\text{in}}$, $Y_{\text{in}}$), and the quantum Langevin equations~(\ref{fluceqcd}) with the cold-damping feedback can be solved in the frequency domain. Based on the steady-state solution, we can calculate the spectra of the position and momentum operators for $N$ mechanical resonators, and then the final mean phonon numbers in these resonators can be obtained by integrating the corresponding fluctuation spectra.

\subsection{Final mean phonon numbers \label{sec3C}}

Mathematically, the final mean phonon numbers in $N$ mechanical resonators can be obtained by the relation~\cite{Genes2008PRA,Lai2018PRA}
\begin{equation}
n^{f}_{j\in[1,N]}=\frac{1}{2}[\langle\delta q_{j}^{2}\rangle +\langle\delta p_{j}^{2}\rangle-1],\label{finalphonumber}
\end{equation}
where $\langle\delta q_{j}^{2}\rangle$ and $\langle\delta p_{j}^{2}\rangle$ are, respectively, the variances of the position and momentum operators. These variances can be obtained by solving Eq.~(\ref{fluceqcd}) in the frequency domain, and integrating the corresponding fluctuation spectra,
\begin{subequations}
\label{specintegral}
\begin{align}
\langle\delta q_{j\in[1,N]}^{2}\rangle=&\frac{1}{2\pi}\int_{-\infty}^{\infty}S_{q_{j}}(\omega)d\omega,\\
\langle\delta p_{j\in[1,N]}^{2}\rangle=&\frac{1}{2\pi\omega^{2}_{j}}\int_{-\infty}^{\infty}\omega^{2}S_{q_{j}}(\omega)d\omega.
\end{align}
\end{subequations}
Here, the fluctuation spectra of the position and momentum operators for the corresponding resonators are defined by
\begin{equation}
S_{o}(\omega)=\int_{-\infty}^{\infty}e^{-i\omega\tau}\langle \delta o(t+\tau) \delta o(t)\rangle_{\textrm{ss}}d\tau,\hspace{0.5 cm}(o=q_{j},p_{j}), \label{spectrumtimedomain}
\end{equation}
where $\langle \cdot\rangle_{\textrm{ss}}$ denotes the steady-state average of the system. The fluctuation spectrum can also be expressed in the frequency domain as
\begin{equation}
\langle\delta\tilde{o}(\omega)\delta\tilde{o}(\omega')\rangle_{\textrm{ss}}=S_{o}(\omega) \delta(\omega+\omega'), \hspace{0.5 cm}(o=q_{j},p_{j}).\label{spectrumfdomain}
\end{equation}
Below, we will solve this system in the frequency domain.

\section{Cooling of a two-mechanical-resonator chain\label{sec4}}

In this section, we study the cooling of a two-mechanical-resonator chain by analyzing the effective susceptibilities and noise spectra. We also find the laser-cooling rates of the two mechanical resonators.

\subsection{Analytical results of the effective susceptibilities, cooling rates, and noise spectra \label{sec4A}}

\begin{figure*}[tbp]
\center
\includegraphics[ width=1 \textwidth]{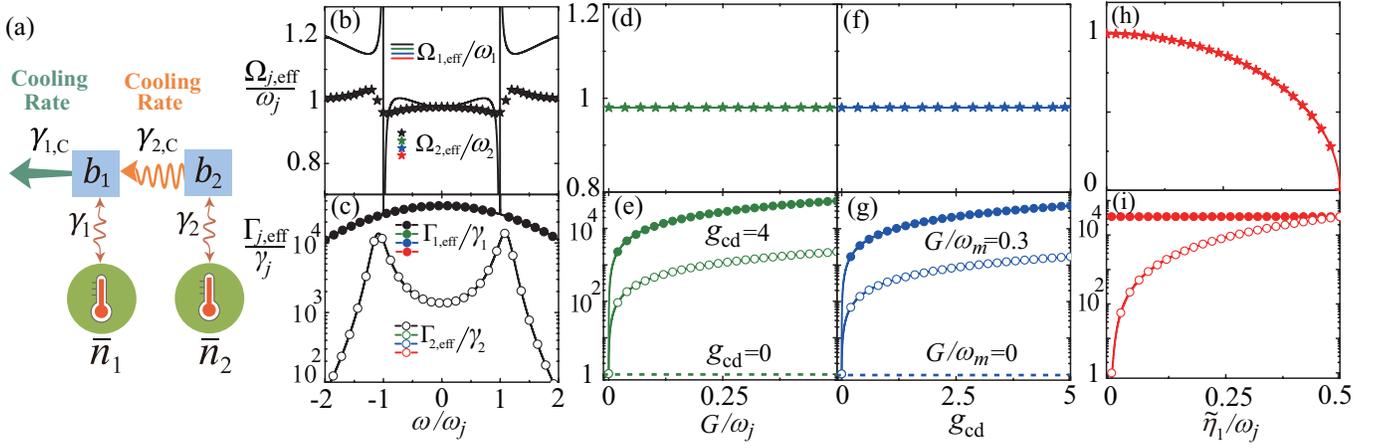}
\caption{(a) Cooling mechanism of a two-mechanical-resonator chain. Here $\gamma_{1,\text{C}}$ and $\gamma_{2,\text{C}}$ denote the cooling rates of the first and second mechanical resonators, respectively; $\bar{n}_{1}$ and $\bar{n}_{2}$ are, respectively, the thermal occupancies of the physical environment for the first and second mechanical resonators, with coupling rates $\gamma_{1}$ and $\gamma_{2}$. (b) The effective mechanical frequencies $\Omega _{j,\text{eff}}(\omega)$ [see Eqs.~(\ref{effective}a) and ~(\ref{effective}b)], and (c) the effective mechanical dampings $\Gamma _{j,\text{eff}}(\omega)$ [see Eq.~(\ref{effective}c)] versus frequency $\omega$ when the effective optomechanical coupling strength $G/\omega_{m}=0.3$, feedback gain $g_{cd}=4$, and effective nearest-neighbor coupling $\tilde{\eta}_{1}/\omega_{m}=0.1$. (d) $\Omega _{j,\text{eff}}(\omega)$ and (e) $\Gamma_{j,\text{eff}}(\omega)$ versus $G$ when $\omega=0$, $g_{cd}=4$, and $\tilde{\eta}_{1}/\omega_{m}=0.1$. The green dashed line at the bottom denotes the case when $g_{cd}=0$. (f) $\Omega _{j,\text{eff}}(\omega)$ and (g) $\Gamma _{j,\text{eff}}(\omega)$ versus $g_{cd}$ when $\omega=0$, $G/\omega_{m}=0.3$, and $\tilde{\eta}_{1}/\omega_{m}=0.1$. The blue dashed line at the bottom denotes the case of $G/\omega_{m}=0$. (h) $\Omega _{j,\text{eff}}(\omega)$ and (i) $\Gamma _{j,\text{eff}}(\omega)$ vs $\tilde{\eta}_{1}$ when $\omega=0$, $G/\omega_{m}=0.3$, and $g_{cd}=4$. The parameters used are given by $\omega_{j=1,2}=\omega_{m}=2\pi\times10$ MHz, $\gamma_{j=1,2}/\omega_{m}=10^{-5}$, $\omega_{\text{fb}}/\omega_{m}=3$, and $\kappa/\omega_{m}=3.5$. }
\label{weffreff}
\end{figure*}

In the two-mechanical-resonator case, the position fluctuation spectra of the two mechanical resonators can be obtained as
\begin{subequations}
\label{spectra12}
\begin{eqnarray}
S_{q_{1}}(\omega )&=&|\chi _{1,\text{eff}}(\omega )|^{2}\Big[S_{\text{fb},1}(\omega )+S_{\text{rp},1}(\omega )\notag \\
&&+S_{\text{th},1}(\omega )+S_{\text{me},1}(\omega )\Big], \\
S_{q_{2}}(\omega )&=&|\chi _{2,\text{eff}}(\omega )|^{2}\Big[S_{\text{th},2}(\omega )+S_{\text{me},2}(\omega )\Big].
\end{eqnarray}
\end{subequations}
Here we introduce the \emph{effective susceptibility} of the $j$th ($j=1,2$) mechanical resonator as
\begin{equation}
\chi_{j,\text{eff}}(\omega )=\omega _{j}[\Omega _{j,\text{eff}}^{2}(\omega )-\omega ^{2}-i\omega \Gamma _{j,\text{eff}}(\omega )]^{-1},\label{susceptibility}
\end{equation}
where $\Omega _{j,\text{eff}}(\omega )$ and $\Gamma _{j,\text{eff}}(\omega )$ are, respectively, the \emph{effective damping rate and resonance frequency} of the $j$th mechanical resonator, defined as
\begin{subequations}
\label{effective}
\begin{eqnarray}
\Omega _{1,\text{eff}}(\omega ) &=&\Bigg[\omega _{1}^{2}+\frac{Gg_{\text{cd}%
}\omega ^{2}\omega _{\text{fb}}\omega _{1}(\kappa +\omega _{\text{fb}})}{%
(\kappa ^{2}+\omega ^{2})(\omega ^{2}+\omega _{\text{fb}}^{2})}  \notag \\
&&+\frac{4\tilde{\eta}_{1}^{2}\omega _{1}\omega _{2}(\omega ^{2}-\omega
_{2}^{2})}{\gamma _{2}^{2}\omega ^{2}+( \omega ^{2}-\omega
_{2}^{2}) ^{2}}\Bigg]^{1/2}, \\
\Omega _{2,\text{eff}}(\omega ) &=&\Bigg[\omega _{2}^{2}+\frac{4\tilde{\eta}%
_{1}^{2}A(\omega )}{C(\omega )}\Bigg]^{1/2}, \\
\Gamma _{j,\text{eff}}(\omega ) &=&\gamma _{j}+\gamma_{j,\text{C}}(\omega).
\end{eqnarray}
\end{subequations}
In Eq.~(\ref{effective}c), the \emph{cooling rates} of the first and second mechanical resonators are defined as
\begin{subequations}
\label{coolingrate0}
\begin{eqnarray}
\gamma_{1,\text{C}}(\omega)&=&\frac{Gg_{\text{cd}}\omega _{%
\text{fb}}\omega _{1}(\kappa \omega _{\text{fb}}-\omega ^{2})}{(\kappa
^{2}+\omega ^{2})(\omega ^{2}+\omega _{\text{fb}}^{2})}\notag \\
&&+\frac{4\tilde{\eta}_{1}^{2}\omega _{1}\omega _{2}\gamma _{2}}{\gamma
_{2}^{2}\omega ^{2}+( \omega ^{2}-\omega _{2}^{2}) ^{2}}, \\
\gamma_{2,\text{C}}(\omega)&=&\frac{4\tilde{\eta}%
_{1}^{2}B(\omega )}{C(\omega )},
\end{eqnarray}
\end{subequations}
with
\begin{subequations}
\label{coeff0}
\begin{align}
A(\omega ) =&\omega _{1}\omega _{2}[\omega ^{6}-Gg_{\text{cd}}\kappa \omega
^{2}\omega _{1}\omega _{\text{fb}}-\omega ^{2}\omega _{1}(Gg_{\text{cd}}+\omega _{1})\omega _{\text{fb}}^{2}\notag \\
&+\kappa ^{2}(\omega ^{2}-\omega
_{1}^{2})(\omega ^{2}+\omega _{\text{fb}}^{2})+\omega ^{4}(\omega _{\text{fb}}^{2}-\omega _{1}^{2})], \\
B(\omega ) =&\omega _{1}\omega _{2}\{Gg_{\text{cd}}\kappa \omega _{1}\omega_{\text{fb}}^{2}+\kappa ^{2}\gamma _{1}(\omega ^{2}+\omega _{\text{fb}}^{2})\notag \\
&+\omega ^{2}[-Gg_{\text{cd}}\omega _{1}\omega _{\text{fb}}+\gamma
_{1}(\omega ^{2}+\omega _{\text{fb}}^{2})]\}, \\
C(\omega ) =&\{\omega ^{2}(-\kappa \gamma _{1}+\omega ^{2}-\omega_{1}^{2})-[(\kappa +\gamma _{1})\omega ^{2}-\kappa \omega _{1}^{2}]\omega _{\text{fb}}\}^{2}\notag \\
&+\{\omega \lbrack \gamma _{1}\omega ^{2}+\big( \omega^{2}-\omega _{1}(Gg_{\text{cd}}+\omega _{1})\big) \omega _{\text{fb}}\notag \\
&+\kappa (\omega ^{2}-\omega _{1}^{2}-\gamma _{1}\omega _{\text{fb}})]\}^{2}.
\end{align}
\end{subequations}
In Eq.~(\ref{spectra12}), we introduce the feedback-induced noise spectrum $S_{\text{fb},1}(\omega )$ and the radiation-pressure noise spectrum $S_{\text{rp},1}(\omega )$ for the first mechanical resonator, and the mechanical-coupling-induced noise spectrum $S_{\text{me},j}(\omega )$ and the thermal noise spectrum $S_{\text{th},j}(\omega )$ for the $j$th ($j=1,2$) mechanical resonator,
\begin{eqnarray}
S_{\text{fb},1}(\omega ) &=&\frac{g_{\text{cd}}^{2}\omega _{\text{fb}%
}^{2}\omega ^{2}}{4\kappa \zeta (\omega ^{2}+\omega _{\text{fb}}^{2})}, \label{Spectra0}\\
S_{\text{rp},1}(\omega ) &=&\frac{G^{2}\kappa }{\kappa ^{2}+\omega ^{2}}, \\
S_{\text{th},j}(\omega ) &=&\frac{\gamma _{j}\omega }{\omega _{j}}\coth
\left( \frac{\hbar \omega }{2\kappa _{B}T_{j}}\right),  \\
S_{\text{me},1}(\omega ) &=&\frac{4\tilde{\eta}_{1}^{2}\omega _{2}^{2}}{%
\gamma _{2}^{2}\omega ^{2}+\left( \omega ^{2}-\omega _{2}^{2}\right) ^{2}}%
\frac{\gamma _{2}\omega }{\omega _{2}}\coth \left( \frac{\hbar \omega }{%
2\kappa _{B}T_{2}}\right),  \\
S_{\text{me},2}(\omega ) &=&\frac{\tilde{\eta}_{1}^{2}E(\omega )}{\left\vert
D(\omega )\right\vert ^{2}},\label{Spectra1}
\end{eqnarray}
where we introduce
\begin{subequations}
\label{coeff}
\begin{align}
D(\omega ) =&(\kappa -i\omega )(-i\gamma _{1}\omega -\omega ^{2}+\omega_{1}^{2})\omega +[(\kappa -i\omega )(\gamma _{1}-i\omega )\omega\notag \\
&+Gg_{\text{cd}}\omega \omega _{1}+(i\kappa +\omega )\omega _{1}^{2}]\omega _{\text{fb}},\\
E(\omega ) =&4(\omega ^{2}+\omega _{\text{fb}}^{2})\left[G^{2}\kappa +\frac{\gamma _{1}\omega }{\omega _{1}}\coth \left( \frac{\hbar \omega }{2\kappa_{B}T_{1}}\right) (\kappa ^{2}+\omega ^{2})\right]\omega _{1}^{2}\notag \\
&+\frac{g_{\text{cd}}^{2}\omega ^{2}\omega _{1}^{2}\omega _{\text{fb}}^{2}}{\kappa \zeta }(\kappa ^{2}+\omega ^{2}).
\end{align}
\end{subequations}

We note that the \emph{exact analytical results of the final mean phonon numbers} are obtained based on Eqs.~(\ref{finalphonumber}), (\ref{specintegral}), and (\ref{spectra12}), and these results are presented in the Appendix.

\subsection{Analyses of the effective susceptibilities, laser-cooling rates, and noise spectra\label{sec4AB}}

In the above subsection, we have derived the effective mechanical resonance frequency $\Omega _{j,\text{eff}}$ and damping rate $\Gamma_{j,\text{eff}}$ of the $j$th mechanical resonator [see Eq.~(\ref{effective})]. We have also found the analytical expressions of the final thermal excitations in these mechanical resonators [see Eq.~(\ref{exactcoolresult})]. Now, we study how the feedback loop affects the cooling performance by analyzing the dependence of the mechanical resonance frequency $\Omega _{j,\text{eff}}$ and decay rate $\Gamma_{j,\text{eff}}$ on the loop coupling parameters.

Concretely, Figure~\ref{weffreff} plots the effective mechanical resonance frequencies $\Omega _{j,\text{eff}}$ and decay rates $\Gamma_{j,\text{eff}}$ as functions of the frequency $\omega$, optomechanical coupling $G$, feedback gain $g_{\text{cd}}$, and nearest-neighbor interaction $\tilde{\eta}_{1}$. We can see from Figs.~\ref{weffreff}(b) and \ref{weffreff}(c) that at resonance $\omega=0$, the mechanical frequencies change slightly [$\Omega _{j,\text{eff}}(0)\approx0.98\omega_{m}$], while the effective mechanical dampings are significantly increased [$\Gamma_{1,\text{eff}}(0)\approx3.5\times10^{4}\gamma_{m}$, $\Gamma_{2,\text{eff}}(0)\approx1.5\times10^{3}\gamma_{m}$]. This giant enhancement of the mechanical damping plays an important role in the cooling process for the two mechanical resonators.

We see from Figs.~\ref{weffreff}(e,g) and Eqs.~(\ref{effective},\ref{coolingrate0}) that, when we turn off the optomechanical coupling ($G=0$) or the feedback ($g_{\text{cd}}=0$), these mechanical resonators are uncooled ($\Gamma_{j,\text{eff}}/\gamma_{j}\approx1$, i.e., $\gamma_{j,\text{C}}\ll\gamma_{j}$), i.e., the breaking of the feedback loop ($G=0$ or $g_{\text{cd}}=0$) leads to no actual cooling for the mechanical-resonator chain. This is because the feedback loop applied on the first mechanical resonator acts as a cooling impetus of this mechanical-resonator chain. Moreover, by increasing the optomechanical coupling $G$ or the feedback gain $g_{\text{cd}}$, the effective mechanical decay rates $\Gamma_{j,\text{eff}}$ are exponentially increased [see Figs.~\ref{weffreff}(e) and \ref{weffreff}(g)] while the effective mechanical frequencies $\Omega _{j,\text{eff}}$ are nearly unchanged [see Figs.~\ref{weffreff}(d) and \ref{weffreff}(f)]. For example, the effective mechanical damping of the first mechanical resonator is increased from $\Gamma_{1,\text{eff}}/\gamma_{1}=1$ to values larger than $10^{4}$, and that of the second one is increased from $\Gamma_{2,\text{eff}}/\gamma_{2}=1$ to values larger than $10^{3}$. Physically, increasing the optomechanical coupling $G$ or the feedback gain $g_{\text{cd}}$ enhances the feedback loop, and then substantially enhances the cooling efficiencies of these mechanical resonators.

In the absence of the nearest-neighbor coupling ($\tilde{\eta}_{1}=0$) between the two mechanical resonators, the first mechanical resonator is substantially modulated ($\Gamma_{1,\text{eff}}\approx3.5\times10^{4}\gamma_{1}$) by the feedback loop, while the second one becomes a dissipative harmonic resonator ($\Omega _{2,\text{eff}}=\omega_{2}$, $\Gamma_{2,\text{eff}}=\gamma_{2}$), as shown in Figs.~\ref{weffreff}(h) and~\ref{weffreff}(i). This means that the cooling is feasible for the first mechanical resonator but not for the second one due to a zero-value cooling rate, i.e., $\gamma_{2,\text{C}}=0$ [see Eq.~(\ref{coolingrate0}b)]. Increasing the nearest-neighbor coupling $\tilde{\eta}_{1}$, the effective mechanical frequency $\Omega _{j,\text{eff}}$ decreases, and the effective mechanical damping of the second mechanical resonator significantly increases from $\Gamma_{2,\text{eff}}/\gamma_{2}=1$ to $3.5\times10^{4}$ [see Figs.~\ref{weffreff}(h) and~\ref{weffreff}(i)].

\begin{figure}[tbp]
\center
\includegraphics[width=0.48 \textwidth]{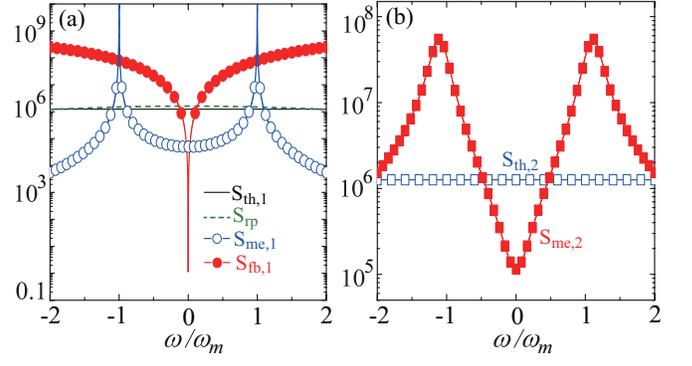}
\caption{The noise spectra of (a) the first and (b) the second mechanical resonators are plotted as functions of the frequency $\omega$. Other parameters are the same as those used in Fig.~\ref{weffreff}.}
\label{spectra}
\end{figure}

To analyze the cooling rates of the two mechanical resonators, we consider the case $\omega=0$ and reexpress Eq.~(\ref{effective}) as
\begin{subequations}
\label{effe}
\begin{eqnarray}
\Omega_{1,\text{eff}} &=&\sqrt{\omega _{1}^{2}-\frac{4\tilde{\eta}_{1}^{2}\omega _{1}}{\omega _{2}}}, \\
\Omega_{2,\text{eff}} &=&\sqrt{\omega _{2}^{2}-\frac{4\tilde{\eta}%
_{1}^{2}\omega _{2}}{\omega _{1}}}, \\
\Gamma_{j,\text{eff}} &=&\gamma _{j}+\gamma _{j,\text{C}},
\end{eqnarray}
\end{subequations}
where $\gamma_{j,\text{C}}$ denotes the cooling rate of the $j$th mechanical resonator, defined as
\begin{subequations}
\label{coolingrate}
\begin{eqnarray}
\gamma_{1,\text{C}} &=&\frac{Gg_{\text{cd}}\omega _{1}}{\kappa }+\frac{4\tilde{\eta}_{1}^{2}\omega _{1}\gamma _{2}}{\omega _{2}^{3}},  \\
\gamma_{2,\text{C}} &=&\frac{4\tilde{\eta}_{1}^{2}\omega _{2}(Gg_{\text{cd}}\omega _{1}+\kappa \gamma _{1})}{\omega _{1}^{3}\kappa }.
\end{eqnarray}
\end{subequations}
We can see from Eqs.~(\ref{effe}a) and (\ref{effe}b) that the effective mechanical frequencies $\Omega_{1,\text{eff}}$ and $\Omega_{2,\text{eff}}$ are modulated only by the nearest-neighbor coupling $\tilde{\eta}_{1}$ between the adjacent resonators. This feature can well explain the phenomenon that the effective mechanical frequencies $\Omega_{j,\text{eff}}$ are \emph{independent of the feedback loop} ($G$ and $g_{\text{cd}}$) but are \emph{sensitive to the nearest-neighbor coupling} $\tilde{\eta}_{1}$ [see Figs.~\ref{weffreff}(d), \ref{weffreff}(f), and \ref{weffreff}(h)].

The parameters $\gamma_{1,\text{C}}$ and $\gamma_{2,\text{C}}$ defined in Eqs.~(\ref{coolingrate}a) and (\ref{coolingrate}b) are, respectively, the feedback-loop and mechanical cooling rates. Here, the feedback-loop cooling rate $\gamma_{1,\text{C}}$ is mainly governed by the radiation pressure $G$ and feedback $g_{\text{cd}}$, while the mechanical cooling rate $\gamma_{2,\text{C}}$ is decided by the mechanical coupling $\tilde{\eta}_{1}$. When we turn off the feedback loop (i.e., $G=0$ or $g_{\text{cd}}=0$), the cooling rates of the two mechanical resonators shown in Eq.~(\ref{coolingrate}) become
\begin{subequations}
\label{coolingrate2}
\begin{eqnarray}
\gamma_{1,\text{C}} &=&\frac{4\tilde{\eta}_{1}^{2}\omega_{1}\gamma _{2}}{\omega _{2}^{3}},  \\
\gamma_{2,\text{C}} &=&\frac{4\tilde{\eta}_{1}^{2}\omega_{2}\gamma _{1}}{\omega _{1}^{3}}.
\end{eqnarray}
\end{subequations}
We can see from Eqs.~(\ref{coolingrate2}a) and (\ref{coolingrate2}b) that, when the feedback loop is broken ($G=0$ or $g_{\text{cd}}=0$), the cooling rates of the two mechanical resonators are largely suppressed due to $\gamma_{j,\text{C}}\ll \gamma_{j}$, as shown in Figs.~\ref{weffreff}(e) and~\ref{weffreff}(g). Physically, to realize the ground-state cooling of this mechanical-resonator chain, the cooling rate $\gamma_{j,\text{C}}$ should be larger than the thermal-reservoir coupling rate $\gamma_{j}$ (i.e., $\gamma_{j,\text{C}}\gg\gamma_{j}$), and the cooling rate of the first resonator should be much larger than that of the second one ($\gamma_{1,\text{C}}\gg\gamma_{2,\text{C}}$). These results coincide with those shown in Figs.~\ref{weffreff}(c), \ref{weffreff}(e), \ref{weffreff}(g), and \ref{weffreff}(i). Thus, the thermal excitations stored in the second mechanical resonator can be extracted into the first one by the cascade cooling channel $\gamma_{2,\text{C}}$.

In fact, the cooling of the two mechanical resonators can be explained based on the noise spectra [see Eqs.~(\ref{Spectra0}-\ref{Spectra1})] of the resonators. In Fig.~\ref{spectra}, we plot the noise spectra of the two mechanical resonators as functions of the frequency $\omega$. For the first mechanical resonator, we find that at $\omega=0$, the contribution from the feedback noise $S_{\text{fb},1}(\omega )$ is much smaller than those from the thermal noise $S_{\text{th},1}(\omega )$, the radiation-pressure noise $S_{\text{rp}}(\omega )$, and the mechanical-coupling noise $S_{\text{me},1}(\omega )$, as shown in Fig.~\ref{spectra}(a). For the second mechanical resonator, the mechanical-coupling-noise contribution is much less than that of the thermal noise when $\omega=0$, i.e., $S_{\text{me},2}(0)\ll S_{\text{th},2}(0)$ [see Fig.~\ref{spectra}(b)]. Therefore, the efficient cooling of the two-mechanical-resonator chain can be achieved because the thermal noise stored in these resonators is significantly suppressed by the cold-damping feedback.

\subsection{Ground-state cooling\label{sec4B}}

To investigate the cooling rule of this cascade optomechanical system, we first consider the cooling of a two-mechanical-resonator system. Physically, the first mechanical resonator undergoing cold damping can be directly cooled to its quantum ground-state by the feedback-loop cooling channel ($\gamma_{1,\text{C}}$), and the second one can also experience a cooling process via the cascade-cooling channel ($\gamma_{2,\text{C}}$) between the adjacent mechanical resonators [see Fig.~\ref{weffreff}(a)]. Below, we show in detail the dependence of the cooling performance of the two mechanical resonators on the system parameters.

\begin{figure}[tbp]
\center
\includegraphics[width=0.48 \textwidth]{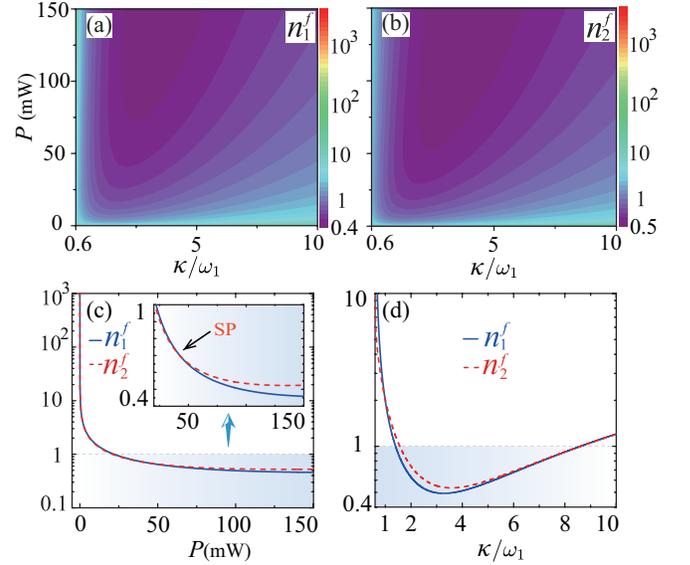}
\caption{The final average phonon numbers (a) $n_{1}^{f}$ and (b) $n_{2}^{f}$ in the two mechanical resonators versus the laser power $P$ and the cavity-field decay rate $\kappa$ for standard feedback. (c) $n^{f}_{1}$ (blue solid curves) and $n^{f}_{2}$ (red dashed curves) as functions of the laser power $P$ when $\kappa/\omega_{1}=3.5$. (d) $n^{f}_{1}$ (blue solid curves) and $n^{f}_{2}$ (red dashed curves) as functions of $\kappa$ when $P=100$ mW. The parameters used are given by $\omega_{1}/2\pi=\omega_{2}/2\pi=10$ MHz, $\omega_{\text{fb}}/\omega_{1}=3$, $g_{\text{cd}}=0.9$, $\zeta=0.8$, $\gamma_{1}/\omega_{1}=\gamma_{2}/\omega_{1}=10^{-5}$, $\omega_{c}/\omega_{1}=2.817\times10^{7}$, $\tilde{\eta}_{1}/\omega_{1}=0.05$, $m_{1}=m_{2}=250$ ng, $\bar{n}_{1}=\bar{n}_{2}=10^{3}$, $L=0.5$ mm, and $\lambda=1064$ nm.}
\label{Pk}
\end{figure}

In Fig.~\ref{Pk}, we plot the final mean phonon numbers $n_{1}^{f}$ and $n_{2}^{f}$ as functions of the laser power $P$ and the cavity-field decay rate $\kappa$. It is shown that the two mechanical resonators can be cooled efficiently ($n^{f}_{1},n^{f}_{2}<1$) in the unresolved-sideband regime $\kappa/\omega_{m}>1$. This indicates that the simultaneous ground-state cooling of the two mechanical resonators is achievable via the cold-damping feedback. In addition, the optimal cooling performances of the two mechanical resonators are $n^{f}_{1}\approx0.5$ and $n^{f}_{2}\approx0.55$ when $P=100$ mW and $\kappa/\omega_{1}=3.5$. To further elucidate this aspect, in Fig.~\ref{Pk}(c) we show the dependence of the cooling efficiencies of the two mechanical resonators on the laser power $P$. We find that when $P<100$ mW, the cooling becomes less efficient when decreasing the laser power $P$. These results indicate that the feedback loop plays the role of a cooling impetus, and that these mechanical resonators cannot be cooled because the feedback loop is broken when $P\rightarrow0$ [see Fig.~\ref{Figmodel}(b)]. Particularly, it shows one switch point (SP) (i.e., the symmetric cooling point $n_{1}^{f}=n_{2}^{f}$) in Fig.~\ref{Pk}(c). This means that a flexible asymmetric-to-symmetric or inverse cooling switch can be achieved by appropriately engineering the laser power $P$. Furthermore, we can see from Fig.~\ref{Pk}(d) that the optimal cooling of the two resonators is achieved around $\kappa/\omega_{1}=3$. This point is different from that in the sideband cooling method, in which the optimal cooling is reached in the resolved-sideband regime~\cite{Wilson-Rae2007PRL,Marquardt2007PRL,Teufel2011Nature}.

\begin{figure}[tbp]
\center
\includegraphics[ width=0.48 \textwidth]{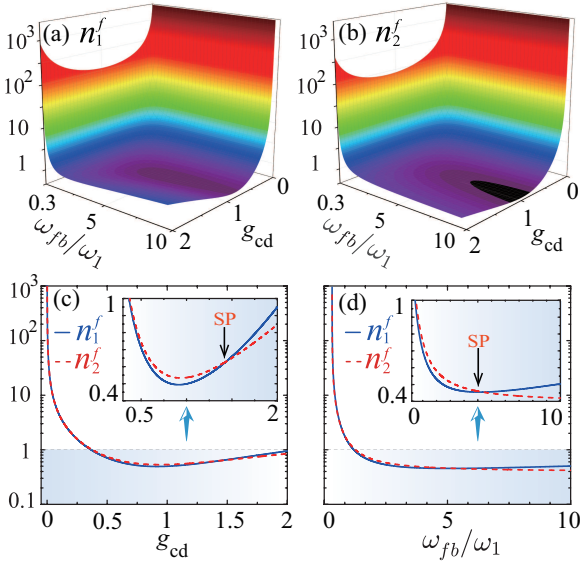}
\caption{The final average phonon numbers (a) $n_{1}^{f}$ and (b) $n_{2}^{f}$ versus the feedback gain $g_{\text{cd}}$ and the feedback bandwidth $\omega_{\text{fb}}$. (c) $n^{f}_{1}$ (blue solid curves) and $n^{f}_{2}$ (red dashed curves) as functions of the feedback gain $g_{\text{cd}}$ when $\omega_{\text{fb}}/\omega_{1}=3$. (d) $n^{f}_{1}$ (blue solid curves) and $n^{f}_{2}$ (red dashed curves) as functions of the feedback bandwidth $\omega_{\text{fb}}$ when $g_{\text{cd}}=0.9$. Here we consider these parameters $P=100$ mW and $\kappa/\omega_{1}=3.5$. Other parameters used are the same as those used in Fig.~\ref{Pk}.}
\label{wfbgcd}
\end{figure}

In Fig.~\ref{wfbgcd}, we investigate the dependence of the cooling efficiencies of the two mechanical resonators on the feedback gain $g_{\text{cd}}$ and the feedback bandwidth $\omega_{fb}$. We find that the optimal cooling can be achieved for the parameters $g_{\text{cd}}>0.5$ and $\omega_{fb}/\omega_{m}>2$. However, when $g_{\text{cd}}\rightarrow0$, the two mechanical resonators are uncooled due to the breaking of the feedback loop, as shown in Fig.~\ref{wfbgcd}(c). In the absence of the feedback loop (i.e., $G=0$ or $g_{\text{cd}}=0$), we can see from Eqs.~(\ref{coolingrate}a) and (\ref{coolingrate}b) that the cooling rates of these resonators are largely suppressed owing to $\gamma_{j,\text{C}}\ll \gamma_{j}$.

When the feedback bandwidth $\omega_{fb}\rightarrow0$, the two mechanical resonators cannot be cooled, as shown in Fig.~\ref{wfbgcd}(d). This is because a lower feedback bandwidth indicates a longer time delay of the feedback loop, and this leads to a lower cooling efficiency in this system. In addition, there is one SP in Figs.~\ref{wfbgcd}(c) and ~\ref{wfbgcd}(d), respectively. These results indicate that by appropriately engineering the laser power $P_{L}$ or the feedback $\omega_{\text{fb}}$ ($g_{\text{cd}}$), a flexible cooling switch between symmetrical and asymmetrical ground-state cooling of these mechanical resonators can be realized.

The feedback loop provides a direct cooling channel ($\gamma_{1,\text{C}}$) to extract the thermal excitations in the first mechanical resonator, and then the second resonator can be cooled by the mechanical cooling channel ($\gamma_{2,\text{C}}$) between the two mechanical resonators [see Fig.~\ref{weffreff}(a)]. Consequently, the optimal cooling of the first mechanical resonator plays a key role on that of the second one. This is because the cooling efficiency of the second mechanical resonator depends on the rotating-wave coupling between the two mechanical resonators. This coupling is determined by both the resonance frequencies of the two mechanical resonators and the coupling strength between them.

\begin{figure}[tbp]
\center
\includegraphics[ width=0.48 \textwidth]{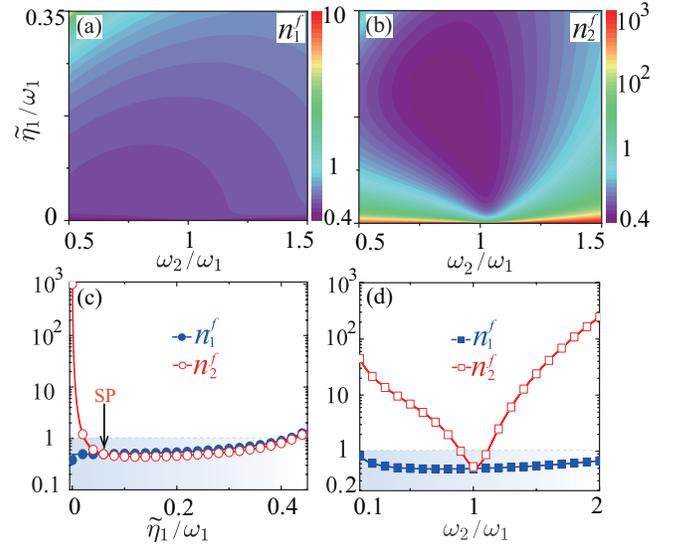}
\caption{The final average phonon numbers (a) $n_{1}^{f}$ and (b) $n_{2}^{f}$ versus the mechanical coupling $\tilde{\eta}_{1}$ and the frequency ratio $\omega_{2}/\omega_{1}$. (c,d) $n^{f}_{1}$ (blue) and $n^{f}_{2}$ (red) as functions of $\tilde{\eta}_{1}/\omega_{1}$ when $\omega_{1}=\omega_{2}$. (e) $n^{f}_{1}$ (blue) and $n^{f}_{2}$ (red) versus the ratio $\omega_{2}/\omega_{1}$ when $\tilde{\eta}_{1}/\omega_{1}=0.05$. Here we choose $P=100$ mW and $\kappa/\omega_{1}=3.5$. Other parameters are the same as those given in Fig.~\ref{Pk}.}
\label{etaW2}
\end{figure}

To further elucidate this effect, the final mean phonon numbers $n_{1}^{f}$ and $n_{2}^{f}$ are plotted as functions of the mechanical coupling strength $\tilde{\eta}_{1}$ and the frequency ratio $\omega_{2}/\omega_{1}$, as shown in Figs.~\ref{etaW2}(a) and~\ref{etaW2}(b). We find that the two mechanical resonators can be simultaneously cooled to their quantum ground states within a large mechanical frequency bandwidth, and that the optimal cooling is located at $\omega_{2}/\omega_{1}\approx1$. The mechanical coupling between the two resonators provides a mechanical cooling channel ($\gamma_{2,\text{C}}$) for the second resonator. This point can be confirmed based on no actual
cooling for the second mechanical resonator when $\tilde{\eta}_{1}=0$ [see Fig.~\ref{etaW2}(c)]. In the weak-coupling region $\tilde{\eta}_{1}/\omega_{m}<0.06$, the cooling performance of the first mechanical resonator becomes worse while that of the second one becomes better with increasing $\tilde{\eta}_{1}$, i.e., $n_{1}^{f}<n_{2}^{f}$. The reason for this phenomenon is that the cooling channel of the second resonator is directly provided by the first resonator which is cooled by the feedback loop, while the second resonator will encumber the cooling efficiency of the first resonator. In the region $0.06<\tilde{\eta}_{1}/\omega_{m}<0.45$, the cooling performance of the two resonators shows an opposite result (i.e., $n_{1}^{f}>n_{2}^{f}$) in comparison with that in the region $\tilde{\eta}_{1}/\omega_{m}<0.06$. Physically, with the increase of the mechanical coupling strength, the CRW interaction terms, which simultaneously create phonon excitations in the two resonators, will become more and more important, and then the cooling of the first resonator will be suppressed largely. Moreover, the symmetrical cooling ($n_{1}^{f}=n_{2}^{f}$) of the two mechanical resonators can be achieved when the mechanical coupling strength takes $\tilde{\eta}_{1}/\omega_{m}=0.06$. These results mean that, when the nearest-neighbor coupling strength $\tilde{\eta}_{1}$ takes a proper value ($\tilde{\eta}_{1}/\omega_{m}<0.06$), the cooling efficiency is higher for the mechanical oscillator which is closer to the cavity.

Additionally, we can see from Fig.~\ref{etaW2}(d) that the optimal cooling efficiency of the two mechanical resonators emerges when the two resonators are resonant and near resonant ($\omega_{2}$ around $\omega_{1}$). Physically, the efficiency of energy extraction from the second resonator decreases with increasing this detuning, and the counter rotating-wave interaction becomes important when this detuning becomes comparable to the mechanical frequencies. When $\omega_{2}/\omega_{1}>2$, the cooling channel of the second resonator is almost turned off (i.e., $\gamma_{2,\text{C}}\approx0$), owing to the approximately negligible mechanical interaction under the condition $\tilde{\eta}_{1}/\vert\omega_{1}-\omega_{2}\vert\ll1$. In this case, the second mechanical resonator will be thermalized by its thermal bath, then the system becomes a typical optomechanical system consisting of an optical cavity and a mechanical resonator. These results provide the possibility to reach simultaneous ground-state cooling of both degenerate and nondegenerate mechanical resonators in the unresolved-sideband regime.

\begin{figure}[tbp]
\center
\includegraphics[ width=0.48 \textwidth]{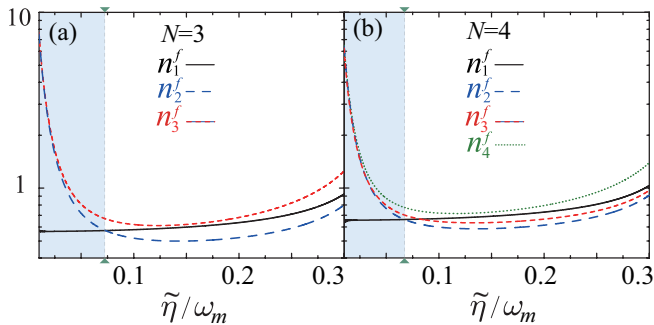}
\caption{The final average phonon numbers $n^{f}_{j}$ in these mechanical resonators are plotted as a function of the mechanical coupling $\tilde{\eta}$ for the cases of (a) $N = 3$ and (b) $N = 4$. We assume that all the mechanical resonators have the same resonant frequencies $\omega_{j}=\omega_{m}$, decay rates $\gamma_{j}=\gamma_{m}$, thermal phonon numbers $\bar{n}_{j}=\bar{n}=10^{3}$, and mechanical coupling strengths $\tilde{\eta}_{j}=\tilde{\eta}$. Other parameters used are the same as those given in Fig.~\ref{Pk}.}
\label{n34}
\end{figure}

\section{Cooling of a coupled $N$-resonator chain\label{sec5}}

We now extend our cold-damping-feedback cooling scheme to the case of an $N$-mechanical-resonator chain. We consider an optical cavity coupled to an array of $N$ mechanical resonators coupled in series, as shown in Figs.~\ref{Figmodel}(a) and ~\ref{Figmodel}(b). The feedback loop is applied on the first mechanical resonator and other nearest-neighboring mechanical resonators are coupled to each other through the mechanical interactions. The first mechanical resonator can be cooled by the feedback loop, and then the thermal excitations in the later mechanical resonator will be extracted by the former via the mechanical cooling channel. As a result, the physical mechanism behind this cooling scheme could be understood as a cascade-cooling process: akin to a domino effect or chain reaction.

Without loss of generality, we consider the identical-resonator case where all the mechanical resonators have the same resonant frequencies $\omega_{j}=\omega_{m}$, decay rates $\gamma_{j}=\gamma_{m}$, thermal phonon numbers $\bar{n}_{j}=\bar{n}$, and mechanical coupling strengths $\tilde{\eta}_{j}=\tilde{\eta}$. Here, we consider the cases of three and four mechanical resonators (i.e., $N=3,4$) in our simulations. In Fig.~\ref{n34}, we plot the final mean phonon numbers in these
mechanical resonators as a function of the mechanical coupling $\tilde{\eta}$ for the cases of (a) $N = 3$ and (b) $N = 4$. We can see that the final mean phonon numbers in these mechanical resonators can be effectively decreased from $10^{3}$ to below $1$. This indicates that the simultaneous cooling of these mechanical resonators can be achieved by using the cold-damping feedback scheme.

Figure~\ref{n34} shows that, when $\tilde{\eta}\ll\omega_{m}$, the final average phonon numbers successively increase from $n^{f}_{1}$ to $n^{f}_{N}$ (see the shadow areas), i.e., the closer to the optomechanical cavity the resonator is, the smaller the final average phonon number in this resonator is. Physically, the thermal excitations in the first resonator is extracted via the feedback cooling channel, and successively, the thermal phonons stored in the next resonator is extracted by the former via the mechanical cooling channel. In this case, the feedback cooling rate should be much larger than the mechanical cooling rates, and thus the cooling efficiency is higher for the mechanical oscillator which is closer to the cavity. In addition, with the increase of $\tilde{\eta}$, we find an anomalous cooling (i.e., the feedback-cooled resonator is not the coldest) (see the blank areas). This phenomenon can also be explained based on the excitations increase caused by the CRW terms.

\section{Discussion and Conclusion\label{sec6}}

Finally, we present some discussions on the understanding of the cooling problems of our system in the mechanical normal-mode representation. In a two-resonator optomechanical system, a cavity-field mode couples to the first mechanical resonator via the radiation-pressure coupling, and the two mechanical resonators are coupled to
each other through the mechanical interaction. After diagonalizing the coupled mechanical resonators, the model can be described by a multi-mode system where the cavity-field mode couples to two mechanical normal modes. However, we should point out that the frequency difference between the two normal modes depends on the coupling strength between the two resonators. Depending on the relation between the frequency difference and the width of the cooling window, there are two different cases~\cite{Lai2020PRARC,Sommer2019PRL}.
(i) When the frequency difference between the two mechanical normal modes is larger than the effective mechanical linewidth, the simultaneous cooling of these mechanical normal modes is accessible because there is no dark mode in this system~\cite{Lai2020PRARC,Sommer2019PRL}.
(ii) When the frequency difference is smaller than the effective mechanical linewidth, the cooling of the two mechanical normal modes is suppressed, because the dark-mode effect works in the near-degenerate-resonator case. The cooling of these normal modes is less efficient and depends on the number of normal modes. In this case, the final average phonon numbers in these mechanical normal modes are $\bar{n}(N-1)/N$ with $\bar{n}_{j}=\bar{n}$~\cite{Lai2020PRARC,Sommer2019PRL}.

In conclusion, we have studied how to realize the simultaneous ground-state cooling of a mechanical-resonator chain coupled to an optomechanical cavity via a standard cold-damping feedback technique. We have found that the entire chain is cooled via a domino effect or chain reaction through the system. We have obtained analytical results for the effective susceptibilities, noise spectra, final mean phonon numbers, and cooling rates of these mechanical resonators. We have also found the optimal-cooling condition for these resonators. In addition, we have found that by appropriately engineering the laser power or the feedback applied on the first mechanical resonator, a flexible switch between symmetric and asymmetric ground-state cooling can be achieved. This could potentially be used to prepare symmetric quantum states in coupled mechanical systems. Our cascade-cooling proposal works for both degenerate and nondegenerate mechanical resonators in the unresolved-sideband regime. This work will pave the way for studying and observing quantum coherence effects involving multiple mechanical modes.

\begin{acknowledgments}
D.-G.L. thanks Dr. Ken Funo and Li Yuan for their useful comments on the manuscript. J.-Q.L. is supported in part by National Natural Science Foundation of China (Grants No. 11822501, No. 11774087, and No. 11935006), Hunan Science and Technology Plan Project (Grant No. 2017XK2018), and the Science and Technology Innovation Program of Hunan Province (Grant No. 2020RC4047). B.-P.H. is supported in part by National Natural Science Foundation of China (Grant No.~11974009). F.N. is supported
in part by: Nippon Telegraph and Telephone Corporation (NTT) Research, the Japan Science and Technology Agency (JST) [via the Quantum Leap Flagship Program (Q-LEAP) program, the Moonshot RD Grant Number JPMJMS2061, and the Centers of Research Excellence in Science and Technology (CREST) Grant No. JPMJCR1676], the Japan Society for the Promotion of
Science (JSPS) [via the Grants-in-Aid for Scientific Research (KAKENHI) Grant No. JP20H00134 and the JSPS-RFBR Grant No. JPJSBP120194828], the Army Research Office (ARO) (Grant No. W911NF-18-1-0358), the Asian Office of Aerospace Research and Development (AOARD) (via Grant No. FA2386-20-1-4069), and the Foundational Questions Institute Fund (FQXi) via Grant
No. FQXi-IAF19-06.

\end{acknowledgments}

\appendix*
\section{Calculation of the final mean phonon numbers\label{appendixa}}
In this appendix, we present the exact analytical results of the final mean phonon numbers in the two-mechanical-resonator case. As shown in Sec.~\ref{sec3C}, by calculating the integral in Eq.~(\ref{specintegral}) for the position and momentum fluctuation spectra, the exact final phonon numbers in the two mechanical resonators can be obtained as~\cite{Genes2008PRA,Sommer2019PRL}
\begin{equation}
\label{exactcoolresult}
n_{l=1,2}^{f}=\frac{1}{2}\left( \frac{iD_{6}^{(l)}}{2\Delta _{6}}+\frac{%
iM_{6}^{(l)}}{2\Delta _{6}}-1\right) .
\end{equation}
Here, we introduce the variables
\begin{eqnarray}
\Delta _{6}
&=&a_{5}%
\{a_{4}(-a_{1}a_{2}a_{3}+a_{3}^{2}+a_{1}^{2}a_{4})+[-a_{2}a_{3}+a_{1}(a_{2}^{2}-2a_{4})]a_{5}
\notag \\
&&+a_{5}^{2}\}-[a_{3}^{3}-a_{1}a_{3}(a_{2}a_{3}+3a_{5})  \notag \\
&&+a_{1}^{2}(a_{3}a_{4}+2a_{2}a_{5})]a_{6}+a_{1}^{3}a_{6}^{2},
\end{eqnarray}
\begin{eqnarray}
D_{6}^{(l=1,2)}
&=&[-a_{3}a_{4}a_{5}+a_{3}^{2}a_{6}+a_{5}(a_{2}a_{5}-a_{1}a_{6})]b_{1}^{(l)}+(a_{1}a_{4}a_{5}
\notag \\
&&-a_{5}^{2}-a_{1}a_{3}a_{6})b_{2}^{(l)}+(-a_{1}a_{2}a_{5}+a_{3}a_{5}+a_{1}^{2}a_{6})b_{3}^{(l)}
\notag \\
&&+[-a_{3}^{2}-a_{1}^{2}a_{4}+a_{1}(a_{2}a_{3}+a_{5})]b_{4}^{(l)}  \notag \\
&&+\frac{1}{a_{6}}%
[a_{3}^{2}a_{4}-a_{2}a_{3}a_{5}+a_{5}^{2}+a_{1}^{2}(a_{4}^{2}-a_{2}a_{6})
\notag \\
&&+a_{1}(-a_{2}a_{3}a_{4}+a_{2}^{2}a_{5}-2a_{4}a_{5}+a_{3}a_{6})]b_{5}^{(l)},
\end{eqnarray}
and
\begin{eqnarray}
M_{6}^{(l=1,2)} &=&\frac{1}{\omega _{l}^{2}}\{-[a_{5}\left(
-a_{2}a_{3}a_{4}+a_{2}^{2}a_{5}+a_{4}(a_{1}a_{4}-a_{0}a_{5})\right)   \notag
\\
&&+\left(
-a_{1}a_{3}a_{4}+a_{0}a_{3}a_{5}+a_{2}(a_{3}^{2}-2a_{1}a_{5})\right)
a_{6}+a_{1}^{2}a_{6}^{2}]b_{1}^{(l)}  \notag \\
&&+[-a_{3}a_{4}a_{5}+a_{3}^{2}a_{6}+a_{5}(a_{2}a_{5}-a_{1}a_{6})]b_{2}^{(l)}
\notag \\
&&+(a_{1}a_{4}a_{5}-a_{5}^{2}-a_{1}a_{3}a_{6})b_{3}^{(l)}+(-a_{1}a_{2}a_{5}+a_{3}a_{5}
\notag \\
&&+a_{1}^{2}a_{6})b_{4}^{(l)}+[-a_{3}^{2}-a_{1}^{2}a_{4}+a_{1}(a_{2}a_{3}+a_{5})]b_{5}^{(l)}\},
\end{eqnarray}
where the coefficients in the two-mechanical-resonator case are defined by
\begin{widetext}
\begin{eqnarray}
a_{0} &=&i,  \notag \\
a_{1} &=&\kappa +\gamma _{1}+\gamma _{2}+\omega _{\text{fb}},  \notag \\
a_{2} &=&-i[\omega _{1}^{2}+\omega _{2}^{2}+\gamma _{2}\omega _{\text{fb}%
}+\gamma _{1}(\gamma _{2}+\omega _{\text{fb}})+\kappa (\gamma _{1}+\gamma
_{2}+\omega _{\text{fb}})],  \notag \\
a_{3} &=&-\omega _{1}\omega _{\text{fb}}(Gg_{\text{cd}}+\omega _{1})-\omega
_{2}^{2}(\gamma _{1}+\omega _{\text{fb}})-\gamma _{2}(\omega _{1}^{2}+\gamma
_{1}\omega _{\text{fb}})-\kappa \lbrack \omega _{1}^{2}+\omega _{2}^{2}+\gamma _{2}\omega _{\text{%
fb}}+\gamma _{1}(\gamma _{2}+\omega _{\text{fb}})],  \notag \\
a_{4} &=&i\{\gamma _{1}\omega _{2}^{2}\omega _{\text{fb}}+\omega
_{1}^{2}\left( \omega _{2}^{2}+\gamma _{2}\omega _{\text{fb}}\right) +\kappa
\lbrack \omega _{1}^{2}\omega _{\text{fb}}+\omega _{2}^{2}(\gamma
_{1}+\omega _{\text{fb}})+\gamma _{2}(\omega _{1}^{2}+\gamma _{1}\omega _{\text{fb}})]+\omega
_{1}(Gg_{\text{cd}}\gamma _{2}\omega _{\text{fb}}-4\omega _{2}\tilde{\eta}%
_{1}^{2})\},  \notag \\
a_{5} &=&\omega _{1}\omega _{2}\omega _{\text{fb}}(Gg_{\text{cd}}\omega
_{2}+\omega _{1}\omega _{2}-4\tilde{\eta}_{1}^{2})+\kappa \lbrack \gamma
_{1}\omega _{2}^{2}\omega _{\text{fb}}+\omega _{1}^{2}(\omega _{2}^{2}+\gamma _{2}\omega _{\text{fb}})-4\omega
_{1}\omega _{2}\tilde{\eta}_{1}^{2}],  \notag \\
a_{6} &=&-i\kappa \omega _{1}\omega _{2}\omega _{\text{fb}}(\omega
_{1}\omega _{2}-4\tilde{\eta}_{1}^{2}),
\end{eqnarray}%
\begin{eqnarray}
b_{0}^{\left( 1\right) } &=&0,  \notag \\
b_{1}^{\left( 1\right) } &=&-\frac{\omega _{1}^{2}}{4\kappa \zeta }[g_{\text{cd}}^{2}\omega _{\text{fb}}^{2}+4\kappa \gamma _{1}\zeta (1+2\bar{n}_{1})],
\notag \\
b_{2}^{\left( 1\right) } &=&-\frac{\omega _{1}^{2}}{4\kappa \zeta }\{g_{\text{cd}}^{2}\omega _{\text{fb}}^{2}\left( \kappa ^{2}+\gamma
_{2}^{2}-2\omega _{2}^{2}\right) +4\kappa \lbrack G^{2}\kappa +\gamma
_{1}(1+2\bar{n}_{1})(\kappa ^{2}+\gamma _{2}^{2}-2\omega _{2}^{2}+\omega _{\text{fb}%
}^{2})]\zeta \},  \notag \\
b_{3}^{\left( 1\right) } &=&-\frac{\omega _{1}^{2}}{4\kappa \zeta }\{g_{\text{cd}}^{2}\omega _{\text{fb}}^{2}\left[ \omega _{2}^{4}+\kappa^{2}(\gamma _{2}^{2}-2\omega _{2}^{2})\right] +4\kappa \lbrack G^{2}\kappa
(\gamma _{2}^{2}-2\omega _{2}^{2}+\omega _{\text{fb}}^{2})\notag \\
&&+(1+2\bar{n}_{1})\gamma _{1}\left( \kappa ^{2}\gamma _{2}^{2}-2\kappa^{2}\omega _{2}^{2}+\omega _{2}^{4}+\left( \kappa ^{2}+\gamma
_{2}^{2}-2\omega _{2}^{2}\right) \omega _{\text{fb}}^{2}\right)+4(1+2\bar{n}_{2})\gamma _{2}\omega _{2}^{2}\tilde{\eta}_{1}^{2}]\zeta \},
\notag \\
b_{4}^{\left( 1\right) } &=&-\frac{\omega _{1}^{2}}{4\zeta }\{g_{\text{cd}}^{2}\kappa \omega _{2}^{4}\omega _{\text{fb}}^{2}+4[G^{2}\kappa (\omega
_{2}^{4}+\gamma _{2}^{2}\omega _{\text{fb}}^{2}-2\omega _{2}^{2}\omega_{\text{fb}}^{2})+(1+2\bar{n}_{1})\gamma _{1}\left( \omega _{2}^{4}\omega _{\text{fb}}^{2}+\kappa ^{2}(\omega _{2}^{4}+\gamma _{2}^{2}\omega_{\text{fb}}^{2}-2\omega _{2}^{2}\omega _{\text{fb}}^{2})\right)   \notag \\
&&+4(1+2\bar{n}_{2})\gamma _{2}\omega _{2}^{2}(\kappa ^{2}+\omega _{\text{fb}}^{2})\tilde{\eta}_{1}^{2}]\zeta \},  \notag \\
b_{5}^{( 1) } &=&-\kappa \omega _{1}^{2}\omega _{2}^{2}\omega _{%
\text{fb}}^{2}\{[G^{2}+\kappa \gamma _{1}( 1+2\bar{n}_{1})
]\omega _{2}^{2}+4\kappa (1+2\bar{n}_{2})\gamma _{2}\tilde{\eta}_{1}^{2}\},
\end{eqnarray}
and
\begin{eqnarray}
b_{0}^{\left( 2\right) } &=&0,  \notag \\
b_{1}^{\left( 2\right) } &=&-(1+2\bar{n}_{2})\gamma _{2}\omega _{2}^{2},
\notag \\
b_{2}^{\left( 2\right) } &=&-(1+2\bar{n}_{2})\gamma _{2}\omega
_{2}^{2}(\kappa ^{2}+\gamma _{1}^{2}-2\omega _{1}^{2}+\omega _{\text{fb}}^{2}),  \notag \\
b_{3}^{\left( 2\right) } &=&-\frac{\omega _{2}^{2}}{\kappa \zeta }\{g_{\text{cd}}^{2}\omega _{1}^{2}\omega _{\text{fb}}^{2}\tilde{\eta}_{1}^{2}-2Gg_{\text{cd}}\kappa (1+2\bar{n}_{2})\gamma _{2}\omega _{1}\omega _{\text{fb}}(\kappa +\gamma _{1}+\omega _{\text{fb}})\zeta   \notag \\
&&+\kappa \lbrack \left( 1+2\bar{n}_{2}\right) \gamma _{2}\left( \omega
_{1}^{4}+\gamma _{1}^{2}\omega _{\text{fb}}^{2}-2\omega _{1}^{2}\omega _{\text{fb}}^{2}+\kappa ^{2}(\gamma _{1}^{2}-2\omega _{1}^{2}+\omega _{\text{fb}}^{2})\right)+4\left( 1+2\bar{n}_{1}\right) \gamma _{1}\omega _{1}^{2}\tilde{\eta}_{1}^{2}]\zeta \},  \notag \\
b_{4}^{\left( 2\right) } &=&-\frac{\omega _{2}^{2}}{\zeta }\{2Gg_{\text{cd}%
}(1+2\bar{n}_{2})\gamma _{2}\omega _{1}\omega _{\text{fb}}[\omega
_{1}^{2}\omega _{\text{fb}}+\kappa (\omega _{1}^{2}+\gamma _{1}\omega _{\text{fb}})]\zeta+[(1+2\bar{n}_{2})\gamma _{2}\left( \omega _{1}^{4}\omega _{\text{fb}%
}^{2}+\kappa ^{2}(\omega _{1}^{4}+\gamma _{1}^{2}\omega _{\text{fb}}^{2}-2\omega _{1}^{2}\omega _{\text{fb}}^{2})\right)   \notag \\
&&+4\omega _{1}^{2}\left( G^{2}\kappa +(1+2\bar{n}_{1})\gamma _{1}(\kappa
^{2}+\omega _{\text{fb}}^{2})\right) \tilde{\eta}_{1}^{2}]\zeta+g_{\text{cd}}^{2}\omega _{1}^{2}\omega _{\text{fb}}^{2}[\kappa \tilde{\eta%
}_{1}^{2}+G^{2}(1+2\bar{n}_{2})\gamma _{2}\zeta ]\},  \notag \\
b_{5}^{\left( 2\right) } &=&-\kappa \omega _{1}^{2}\omega _{2}^{2}\omega_{\text{fb}}^{2}\{\kappa (1+2\bar{n}_{2})\gamma _{2}\omega
_{1}^{2}+4[G^{2}+\kappa (1+2\bar{n}_{1})\gamma _{1}]\tilde{\eta}_{1}^{2}\}.
\end{eqnarray}
\end{widetext}

\end{document}